\documentclass{article}

\usepackage{amssymb}
\usepackage{graphicx}
\usepackage{dcolumn}
\usepackage{bm}
\usepackage{times}
\usepackage{epstopdf} 
\usepackage{amssymb}
\usepackage{amsmath}
\usepackage{mathtools}
\begin{document}

\title{The Quantum State Of The Universe From Deformation Quantization and Classical-Quantum Correlation}

\author{M. Rashki$^1$ and S. jalalzadeh$^{1,2}$\thanks{email: shahram.jalalzadeh@unila.edu.br}\\
 {\small Department Of Physics, Shahid Beheshti University, G. C., Evin, Tehran, 19839, Iran.}\\
 {\small Federal University of Latin-American Integration,}\\
{\small Technological Park of Itaipu PO box 2123, Foz do Igua\c{c}u-PR,  85867-670, Brazil.}}

\maketitle
\begin{abstract}
In this paper we study the quantum cosmology of homogeneous and isotropic cosmology, via the Weyl-Wigner-Groenewold-Moyal formalism of phase space quantization, with perfect fluid as a matter source.  The
corresponding quantum cosmology is described by the Moyal-Wheeler-DeWitt equation which
has exact solutions in Moyal  phase space, resulting in Wigner quasiprobability distribution functions peaking around the
classical paths for large values of scale factor. We show that the Wigner functions of these models are peaked around the non-singular universes  with quantum modified density parameter of radiation.
\end{abstract}


\section{Introduction}
The main prediction of the quantum state of the universe is the emergence of classical universe that is a manifest fact of the observable Universe. Therefore, predicting classical cosmology is a constraint on the theory of the state \cite{Hawking}. In minisuperspace models of cosmology, the classical gravitational field equations are often non-linear and the quantization procedure is not unique. There are infinite number of transformations that recast such equations into forms with different finite number of degree of freedom interpretations. Distinctly an appropriate transformation, state and quantization should be those that have some chance of yielding a classical limit not too far removed the original classical predictions \cite{Tucker}.

{ By canonical quantization of gravity, one finds various defects when
one inspects its content in detail. First, since the canonical Hamiltonian
of gravity is written as a linear combination of the constraints, it annihilates
the physical quantum state. Hence, the time evolution is lost from theory.
Besides a new problems aries when one tries to apply the quantum theory of
gravity to cosmology: when our working tool is the wave function, which should be obtained by Wheeler-DeWitt (WDW) equation or path integral, we need to know how it is possible to construct an adequate wave packet that would peaked around the original classical cosmological model \cite{Monerat}.
 In ordinary circumstances, the wave packet reduction
in the Copenhagen interpretation gives rise to no practical problem as far
as we regard the quantum mechanics as describing the dynamics of an ensemble
of identical systems. However, in cosmology, there is only one Universe as
a system. Therefore, it is not clear how the state of the Universe has any
well-defined wave function. Also, in quantum cosmology the observer itself is
an element of Universe. In the standard interpretation of quantum mechanics,
during an observation, the quantum system must interact with a classical domain (measuring instrument, observer body, ...).    In von Neumann's view, the necessity of a classical domain comes
from the way it solves the measurement problem \cite{RR1}. In a conventional
impulsive measurement, where the coupling interaction between the measured
system and the classical measuring devise is of short duration and strong,
the wave function plus measuring devise splits into many branches which do
not overlap, each one containing the measured system in an eigenstate of the measured observable, and the ``pointer'' of the measuring devise pointing
to the corresponding eigenvalue. However, in the end of the measurement processes,
the observer measures only one eigenvalue and  the immediate repeating of measurement gives the same result.
Therefore, the wave function collapses into an eigenstate of the observable
that is registered and the other branches disappear. But,  a real collapse
of wave function cannot be described by the unitary quantum evolution. Therefore,
the Copenhagen interpretation needs to assume the existence of fundamental
process in a measurement which occur outside the quantum system, in a classical
domain. It is obvious that in quantum cosmology, as a quantum theory of whole
Universe, there is no place for classical domain outside of it. Hence, it
seems that the Copenhagen
interpretation cannot be applied to quantum cosmology or it should be improved
by means of further concepts. One possibility is invoking environmental decoherence \cite{RR2}. However, quantum decoherence is not yet a complete answer to the measurement
problem \cite{RR3}. It dose not explain the apparent collapse after the measurement
is completed, or why all but one of the diagonal
elements of the density matrix become null when
the measurement is finished \cite{Pinto}. Also, in its developments like
the consistence histories approach \cite{RR4} the important role played by
observers is not yet explained \cite{RR5} and it is not clear how to describe
a quantum universe when the background geometry is not classical \cite{PP}. Nevertheless,
there are some alternative solutions to the above quantum cosmological difficulties,
which together with decoherence may solve the measurement problem maintaining
the universality of quantum theory and emergence of classical universe. In
this line we should cite the de Broglie-Bohm interpretation of quantum cosmology
\cite{Bohm},
quantum  Hamilton-Jacobi cosmology \cite{Sh} and deformation quantization  of cosmology \cite{RR6}.
 }

The advantage of deformation quantization  is that it makes quantum cosmology look like
the Hamiltonian formalism of cosmology.  In other words, deformation
quantization can be
viewed as a deformation of the structure of the algebra of classical observables,
rather than a radical change in the nature of the observables. This is done by
avoiding the operator formalism \cite{RR6}. Deformation quantization, which is presented as Weyl-Wigner-Groenewold-Moyal phase space quantization,
describes a quantum system in terms of the classical
$c$-number variables \cite{1,2}. This means that operators are mapped
into the $c$-number functions so that their compositions
could be obtained by the star product that is noncommutative
but associative. Therefore, the observables would be
classical functions of the phase space. Quantum structure is
constructed by replacing pointwise products of classical
observables of the phase space $(x,\Pi)$, by star product \cite{4}. The
product of two smooth functions, say $f=f(x,\Pi)$ and $g=g(x,\Pi)$, on a Poisson-Moyal
manifold is given by
\begin{eqnarray}\label{0-1}
f*g=\sum_{n=0}^\infty (i\hbar)^nC_n(f,g),
\end{eqnarray}
where $\hbar$ plays the role of the deformation parameter. The
first term $C_0(f,g)=fg$ denotes the common pointwise product of $f$ and $g$. Also, the
coefficients $C_n(f,g)$ are bidifferential operators, where
their product is noncommutative \cite{5}. These coefficients
satisfy the following properties
\begin{eqnarray}\label{0-2}
\begin{array}{cc}
C_0(f,g)=fg,\\
C_1(f,g)-C_1(g,f)=\{f,g\},\\
\sum\limits_{i+j=n}C_i(C_j(f,g),h)=\sum\limits_{i+j=n}C_i(f,C_j(g,h)),
\end{array}
\end{eqnarray}
where $\{f,g\}$ denotes the ordinary Poisson bracket. In equations (\ref{0-2}), the first
expression means that in the limit, $\hbar\rightarrow0$, the star product of
$f$ and $g$ agrees with the pointwise products of these two
functions in classical phase space. The second expression shows that at the lowest
order of the deformation parameter, the $*$-commutator $[f,g]_*=f*g-g*f$ tends to the Poisson bracket.
In flat spaces, there is a  special star product which has long been
known. In this case, the components of the Poisson tensor can be considered constant and consequently it is possible to define
the following Moyal star product \cite{1}
\begin{eqnarray}\label{Moyal product}
\begin{array}{cc}
f(x,\Pi)\ast_{\textsc{m}}g(x,\Pi)=f(x,\Pi)\exp(\frac{i\hbar}{2}(\overleftarrow{\partial}_x\overrightarrow{\partial}_{\Pi}-\overleftarrow{\partial}_{\Pi}\overrightarrow{\partial}_{x}))g(x,\Pi)\\
=f(x+\frac{i\hbar}{2}\vec\partial_{\Pi},\Pi-\frac{i\hbar}{2}\vec\partial_x)g(x,\Pi).
\end{array}
 \end{eqnarray}
{The last equality in (\ref{Moyal product}) suggests that for a Moyal
star
product $A*_\textsc{m}B$ of two functions $A$ and $B$ the
Weyl-Wigner correspondence reads
\begin{eqnarray}\label{new1}
(A*_\textsc{m}B)(x,\Pi)=A\left(x+\frac{i\hbar}{2}\partial_\Pi,\Pi-\frac{i\hbar}{2}\partial_x\right)B(x,\Pi),
\end{eqnarray}
where now $A\left(x+\frac{i\hbar}{2}\partial_\Pi,\Pi-\frac{i\hbar}{2}\partial_x\right)$ should be understood as an operator acting on ${\mathcal C}^\infty(phase
\,\,space)$. According to this prescription we have to replace the position and momentum variables in $A$ by pseudo-operators containing position and momentum and their derivatives, that is
\begin{eqnarray}\label{new2}
x\longrightarrow x+\frac{i\hbar}{2}\partial_\Pi,\,\,\,\,\,\,\Pi\longrightarrow
\Pi-\frac{i\hbar}{2}\partial_x,
\end{eqnarray}
instead of the usual correspondence $x\longrightarrow x,\,\,\,\,\,\,\Pi\longrightarrow-
\frac{i\hbar}{2}\partial_x$. These pseudo-operators sometimes carry the name Bopp pseudo-operators. These pseudo-operators act, not on functions defined on
$\mathbb{R}^D$ as ordinary Weyl operators do, but on functions (or distributions) defined on
the noncommutative phase space $\mathbb{R}^D\oplus\mathbb{R}^D$. In fact,
Bopp pseudo-differential operators is a tool of choice
for the study of deformation quantization which it reduces to a Weyl calculus of
a particular type. Also, equation (\ref{new1}) shows that  noncommutative quantum
mechanics can also be reduced to Bopp calculus from an operator point of view.
}
One of the most important components of deformation
quantization is the Wigner quasiprobability distribution function \cite{6}.  In fact, it is a generating function for
all spatial autocorrelation functions of a given quantum
mechanical wave function \cite{9}. The relation of Wigner function with wave function of system $\psi_n(x)$ in a $2D$-dimensional
phase space is
\begin{eqnarray}\label{0-3}
\begin{array}{cc}
W_n(x,\Pi)=
\frac{1}{(2\pi\hbar)^D}\int\psi^*_n(x-\frac{\hbar}{2}y)\psi_n(x+\frac{\hbar}{2}y)e^{-i\frac{\Pi.y}{\hbar}}d^Dy.
\end{array}
\end{eqnarray}
  In this formalism of quantum mechanics, expectations
of observables and transition amplitudes are phase space integrals of $c$-number functions,
weighted by the Wigner function, as in statistical mechanics.

In the next section we will investigate the quantum cosmology of a
flat Friedmann-Lema\^{i}tre-Robertson-Walker (FLRW)
universe, filled with radiation plus dust or cosmic string.
Using Wigner function we will show that the deformed cosmology predicts a
good correlation with the corresponding classical cosmology.
Natural units $c = \hbar = 1$ are used throughout
the paper.
\section{Deformation quantization of FLRW cosmology}
{There exist two fundamentally different approaches to quantization of general relativity: The particle physics programme and the canonical quantum gravity programme. In the first programme, the basic entity is the graviton, the quantum of the gravitational
field. Such a particle is deemed to propagate in a background Minkowski spacetime,
$\eta$, and like all elementary particles, is associated with a specific representation of
the Poincar\'e group. By fixing the background
topology and differential structure of spacetime manifold to be that of Minkowski
spacetime, and then splinting spacetime into background, $\eta_{\mu\nu}$, and dynamical, $h_{\mu\nu}$, parts, the field $h_{\mu\nu}$ is quantized by the standard field methods under the
assumption that the gravitational interaction, like other standard matter interactions, involves the exchange of gravitons. On the other hand, the canonical approach to quantum gravity starts with a reference foliation of
spacetime with respect to which the appropriate canonical variables are defined.
In this sense, quantization of gravitation is quantization of the metrical structure of spacetime, which satisfies a dynamical principle and dynamical equations. The essence of this approach is expressed
by the name quantum geometrodynamics. To study the nature of the difficulties of quantum geometrodynamics, it is convenient to use a particular minisuperspace
approximation. The model nature of these studies is a consequence of various
factors, of which the main are: 1) a minisuperspace of definite symmetry  is selected and it is
assumed that the symmetry is preserved in the quantization process. 2) to maintain a definite symmetry of minisuperspace, one requires a material source (perfect fluid,
Yang-Mills fields, scalar fields or spinor fields), which is also quantized when gravitation is quantized. It is clear that in the
domain  of quantum geometrodynamics (Planck scale) the description of the material source
will differ from the one adopted in quantum field theory, and therefore in quantum geometrodynamics the source is
taken into account only formally by the addition of new degrees of freedom to the equations that describe the
dynamical geometry \cite{Lap}.
Due to the quantum nature of the model,
as a first approximation, the matter content should be described by some sorts of  fields, as done in \cite{CC}. However, general
exact solutions are hard to find at the presence of Yang-Mills, scalar or
spinor fields and the Hilbert space structure
is obscure and it is a subtle matter to recover the notion of a semiclassical time \cite{CC, Ish}.  { In addition, since the Bekenstein Bound tells us that the information content of the very early Universe
is zero,  the only physical variable we have to take into account is the scale
factor of the Universe, and the density and pressure of the matter field. Hence, we only have
to quantize the FLRW universe for a  perfect fluid \cite{Tip}.} Also, perfect
fluid has the advantage of introducing a variable which can naturally be identified with time, leading to a well-defined Hilbert space structure \cite{Shu}. Another  attractive feature of the
phenomenological perfect fluid description of matter degree of freedom is that it allows us to treat the barotropic equation of state which allows us to obtain general exact solutions.}
The line element of spatially flat FLRW universe is
\begin{eqnarray}\label{line frw}
ds^2=-N^2(t)dt^{2}
+ a^{2}(t)\left( dx^2+dy^2+dz^2\right),
\end{eqnarray}
where $N(t)$ is the lapse function and $a(t)$ is the scale factor.
The action functional that consists of a gravitational part
and a matter part when the matter field is considered as a
perfect fluid is given by \cite{Haw}
\begin{eqnarray}\label{action1}
S=\frac{1}{16 \pi G}\int{{R}\sqrt{-{g}}d^{4}x}-\int{\rho\sqrt{-{g}}d^{4}{x}},
\end{eqnarray}
where $g$ is the determinant of spacetime metric and $R$ is the Ricci scalar  and  $\rho=\sum\limits_{i}\rho_{i}$ is the total energy density. We assume a
universe filled with noninteracting  perfect fluids with energy densities of $\rho_{i}=\rho_{0i}(\frac{a}{a_{0}})^{-3(\omega_{i}+1)}$  where $\omega_{i}$ denotes the equation of state parameter of $i$-th component of fluid and $\rho_{i0}$ is the energy density at the measuring epoch. { The action (\ref{action1})
reduce to
\begin{eqnarray}\label{action2}
S=-\frac{3V_3}{8\pi G}\int Na^3\left(\frac{a}{N^2}\left(\frac{da}{dt}\right)^2+\frac{8\pi
G}{3}\sum_i\rho_{0i}\left(\frac{a}{a_0}\right)^{-3(\omega_i+1)}\right)dt,
\end{eqnarray}
where $V_3=\int d^3x$ is the spacial volume of 3-metric.
 Let us rewrite  the action (\ref{action2}) in terms of measurable quantities in cosmology. This will help us to compare the quantum cosmological model with the corresponding classical model. First, we define a new lapse function
by  $\tilde{N}=\frac{N}{x}$. By writing the energy density of various components of fluid in terms of corresponding density parameters, $\Omega_i=\frac{8\pi G\rho_{0i}}{3H_0^2}$ ($H_0$ is the Hubble parameter at the measuring epoch) the energy densities will be $\rho_{i}=\frac{3H^{2}_{0}\Omega_i}{8\pi G}(\frac{a}{a_{0}})^{-3(1+\omega_{i})}$. Also if we use a new dimensionless scale factor defined by $x=\frac{a}{a_0}$ and a new dimensionless time coordinate by $\eta=H_0 t$, the Lagrangian of model in conformal frame  up to a multiplicative constant $\frac{3V_3a_0^3H_0}{4\pi
G}$, will be}
 \begin{eqnarray}\label{lagrangian}
    \mathcal{L}=-\frac{1}{2}\left(\frac{\dot x^2}{\tilde{N}}+\tilde{N}\sum_{i}\Omega_{i}{x}^{1-3\omega_{i}}\right),
 \end{eqnarray}
where over
dot denotes differentiation respect to $\eta$. The conjugate momentum to
the  scale factor, $x$, and the primary constraint are
given by
\begin{eqnarray}\label{conjugate}
\Pi_{x}=\frac{\partial{\mathcal{L}}}{\partial{\dot x}}=-\frac{\dot{x}}{\tilde{N}}, \hspace{.6cm}  {\Pi_{\tilde{N}}}=\frac{\partial{\mathcal{L}}}{\partial{\dot{\tilde{ N}}}}=0.
\end{eqnarray}
Consequently, the Hamiltonian corresponding to
Lagrangian (\ref{lagrangian}) will be
\begin{eqnarray}\label{canonical hamilton}
    {H}=\tilde{N}\left(-\frac{{\Pi_x}^{2}}{2}+\frac{1}{2}\sum_{i}\Omega_{i}{x}^{1-3\omega_{i}}\right).
 \end{eqnarray}
In Hamiltonian (\ref{canonical hamilton}), $\tilde{N}$ is a Lagrange multiplier; therefore,
it enforces the Hamiltonian constraint
\begin{eqnarray}\label{constraint}
 \mathcal{H}=-\frac{{\Pi_x}^{2}}{2}+\frac{1}{2}\sum_{i}\Omega_{i}{x}^{1-3\omega_{i}}=0,
 \end{eqnarray}
where ${\mathcal H}$ denotes the super-Hamiltonian. The super-Hamiltonian at the initial time $\eta_0=t_0H_0$ reduces to well-known relation between density parameters $\sum_i\Omega_i=1$.

The deformation quantization of this simple model is
accomplished straightforwardly by replacing the ordinary
products of the observables in phase space by the Moyal
product. Therefore, Hamiltonian constraint (\ref{constraint}) becomes
the Moyal-Wheeler-DeWitt (MWDW) equation by replacing
the classical Hamiltonian (\ref{constraint}) with its deformed
counterpart \cite{RR6}
\begin{eqnarray}\label{sh2}
{\mathcal H}(x,\Pi_x)*_\mathrm{M}W_n(x,\Pi_x)=0.
\end{eqnarray}
 Since the $*_\mathrm{M}$-product involves exponentials of derivative operators, it may be
evaluated in practice through translation of function arguments (see, Eq.(\ref{new1})). Therefore, the MWDW equation (\ref{sh2}) is equivalent to
\begin{eqnarray}\label{b1}
{\mathcal H}\left(x+\frac{i}{2}\vec\partial_{\Pi_x},\Pi_x-\frac{i}{2}\vec\partial_x\right)W(x,\Pi_x)=0.
\end{eqnarray}

Let us now first investigate the classical-quantum correlation  in a radiation-dust filled universe. In this case, the classical super-Hamiltonian (\ref{constraint}) will be
\begin{eqnarray}\label{sh3}
\Pi_x^2-\Omega_dx-\Omega_r=0,
\end{eqnarray}
where $\Omega_d$ and $\Omega_r$ denote the density parameters of dust and radiation respectively, obeying relation $\Omega_d+\Omega_r=1$
at the measuring epoch, $\eta_0$.
Also, the MWDW equation (\ref{b1}) will be
\begin{eqnarray}\label{sh4}
\left(\left(\Pi_x-\frac{i}{2}\vec \partial_x\right)^2-\Omega_d\left(x+\frac{i}{2}\vec\partial_{\Pi_x}\right)-\Omega_r\right)W=0.
\end{eqnarray}
{ Note that we should order the kinetic term as $\Pi_x^2\rightarrow x^{-\alpha}*\Pi_x*x^\alpha*\Pi_x=\Pi_x^2-i\alpha
x^{-1}*\Pi_x$
, where $\alpha$ takes into account the factor ordering ambiguity. This is equivalent to the  factor ordering in corresponding WDW equation given by $\Pi^2_x\rightarrow-x^{-\alpha}\partial_x(x^\alpha\partial_x)$
\cite{Se}.
In this paper we will consider the choice $\alpha=0$ ordering which is equivalent
to the Laplace-Beltrami operator of conformal frame  in the corresponding WDW equation.} The Wigner function is real, hence by separation the real and imaginary parts of MWDW equation (\ref{sh4}) we obtain two coupled partial
differential equations \begin{eqnarray}\label{sh5}
\begin{array}{cc}
\left(-\frac{1}{4}\partial^2_x-\Omega_dx+\Pi^2_x-\Omega_r\right)W(x,\Pi_x)=0,\\
\\
\left(\Pi_x\partial_x+\frac{\Omega_d}{2}\partial_{\Pi_x}\right)W(x,\Pi_x)=0.
\end{array}
\end{eqnarray}
{ The first equation does not involves the partial derivatives of $\Pi_x$.  However, the second phase space equation enforces a special symmetry on the solutions. The solution of second partial differential equation of (\ref{sh5}) is, $W=f(\Pi_x^2-\Omega_dx)$,
where $f$ denotes a general real function.  With the help of definition of new variable $\zeta=\Pi_x^2-\Omega_dx$ and the relation $\frac{\partial^2W(x,\Pi_x)}{\partial x^2}=\Omega^2_d\frac{d^2f(\zeta)}{d\zeta^2}$
following from the chain rule, the first partial differential equation of (\ref{sh5}) reduces to following second order ordinary differential equation
\begin{eqnarray}\label{rr1}
-\frac{\Omega_d^2}{4}\frac{d^2f(\zeta)}{d\zeta^2}+(\zeta-\Omega_r)f(\zeta)=0.
\end{eqnarray}
 } The finite value solution of this equation is
\begin{eqnarray}\label{sh6}
W(x,\Pi_x)={\mathcal N}Ai\left(\left(\frac{2}{\Omega_d}\right)^\frac{2}{3}(\Pi^2_x-\Omega_dx-\Omega_r)\right),
\end{eqnarray}
where ${\mathcal N}=\frac{1}{2\pi}\left(\frac{2}{\Omega_d}\right)^\frac{2}{3}$ and $Ai(\xi)$ denotes the Airy function of first kind. The locus of extremums of the above Wigner function is the following deformed super-Hamiltonian
\begin{eqnarray}\label{n1}
\Pi_x^2-\Omega_dx-\Omega_r+\left(\frac{\Omega_d}{2}\right)^\frac{2}{3}a_n=0,
\end{eqnarray}
where $a_n$ are the zeroes of derivative Airy functions $\frac{d}{d\xi}Ai(-\xi)|_{\xi=a_n}=0$. See Table (1) for the first several terms of $a_n$ sequences.
 \begin{table}[htb]
\caption{Negatives of zeroes of  $Ai'$ for $n= 1,2,3$.} 
\centering 
\begin{tabular}{l c c rrrrrrr}
\hline\hline 
  n & $a_n$  \\
  \hline
  1 & 1.01879...\\
  2&3.24819...\\
  3&4.82009...
 \end{tabular}
\label{t1}
\end{table}
 Hence, equation (\ref{n1}) presents the most probable cosmological solutions.  These solutions (for various values of $a_n$) are the same as the original classical solution (\ref{sh3}) but with modified value of the density parameter of radiation given by
\begin{eqnarray}\label{n2}
\tilde\Omega_r(n)=\Omega_r-\left(\frac{\Omega_d}{2}\right)^\frac{2}{3}a_n.
\end{eqnarray}
It is obvious that all of this solutions (that we have for various values of $a_n$) are non-singular if
\begin{eqnarray}\label{n3}
\Omega_r\left(\frac{\Omega_d}{2}\right)^\frac{2}{3}<a_1.
\end{eqnarray}
   Fig.(1) shows the Wigner function of model with corresponding classical trajectory in phase space. It is seen that a good correlation exists between
the quantum quasiprobability distribution shown in this figure and the classical trajectory in phase space for large values of scale factor, $x$, where the universe is  dust dominated. The observable difference of classical and quantum cosmology is in the values of density parameter of radiation in super-Hamiltonians (\ref{sh3}) and (\ref{n1}). At the very early times the deformed universe is radiation dominated and singularity free. At the late times the predictions of both theories are the same and there  is an exact correlation between classical and quantum universes.
\begin{figure}[htb]
  \centering
  \includegraphics[width=5cm]{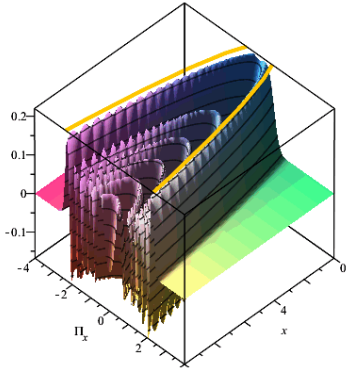}
  \caption{The Wigner function of dust and radiation filled universe. The corresponding classical trajectory is
denoted by goldline loci. This figure is
plotted for  $\Omega_d=\Omega_r=0.5$.}
\end{figure}

Let us now investigate the classical-quantum correlation of a  universe filled with cosmic string perfect fluid, with $\omega_{cs}=-\frac{1}{3}$. In this case, the classical super-Hamiltonian (\ref{constraint}) will be
\begin{eqnarray}\label{sh7}
\Pi^2_x-\Omega_{cs}x^2=0.
\end{eqnarray}
Also, the MWDW equation (\ref{b1}) given by
\begin{eqnarray}\label{sh8}
\left((\Pi_x-\frac{i}{2}\vec\partial_x)^2-\Omega_{cs}(x+\frac{i}{2}\vec\partial_{\Pi_x})^2\right)W=0.
\end{eqnarray}
Separation of real and imaginary parts of the above equation gives two independent equations
\begin{eqnarray}\label{sh9}
\begin{array}{cc}
\left(\Omega_{cs}x\partial_{\Pi_x}+\Pi_x\partial_x\right)W=0,\\
\\
\left(\Pi^2_x-\Omega_{cs}x^2-\frac{1}{4}\partial^2_x+\frac{\Omega_{cs}}{4}\partial^2_{\Pi_x}\right)W=0.
\end{array}
\end{eqnarray}
The finite value solution at the
classical singularity is given by
\begin{eqnarray}\label{sh10}
 W(x,\Pi_x)=\frac{1}{2 \pi\sqrt{\Omega_{cs}} }J_{0}\left(\frac{{\Pi_x}^{2}-\Omega_{cs}{x}^{2}}{\sqrt{\Omega_{cs}}}\right),
\end{eqnarray}
where $J_0(\xi)$ is the Bessel function of order zero. In this very simple model, the locus of extremums of
Wigner function are given by
\begin{eqnarray}\label{re1}
\Pi_x^2-\Omega_{cs}x^2-\sqrt{\Omega_{cs}}j_n=0,
\end{eqnarray}
where $j_n$ are the zeroes of derivative Bessel function. Table (2) shows the first several zeroes of $\frac{dJ_0(\xi)}{d\xi}$.
\begin{table}[htb]
\caption{Zeroes of  Bessel function, $j_n$, for $n= 1,2,3$.} 
\centering 
\begin{tabular}{l c c rrrrrrr}
\hline\hline 
  n & $j_n$  \\
  \hline
  1 & 0\\
  2&3.8317...\\
  3&7.0155...
 \end{tabular}
\label{t2}
\end{table}
The modified super-Hamiltonian (\ref{re1}) represents a universe filled with cosmic string and radiation fluids, where the density parameter of radiation is given by
\begin{eqnarray}\label{re2}
\Omega_r=\sqrt{\Omega_{cs}}j_n.
\end{eqnarray}
Note that the radiation part of (\ref{re1}) has totally quantum origin and the corresponding classical universe is filled only with the cosmic string  fluid. In the late times, where the universe is cosmic string dominated, the predictions of both models are the same. But for very small values of scale factor, the emerged universes for various values of $j_n$ are non-singular and radiation dominated. The first zero of derivative Bessel function is zero, consequently for $j_0=0$ the prediction of quantum cosmology is a cosmic string filled singular universe, same as the corresponding classical universe (\ref{sh7}).
Fig.(2) shows the corresponding Wigner function with corresponding classical trajectory in phase space. For large values of scale factor, $x$, there is a good correlation between
the quantum quasiprobability distribution and the classical trajectory in phase space.
\begin{figure}[htb]
  \centering
  \includegraphics[width=5cm]{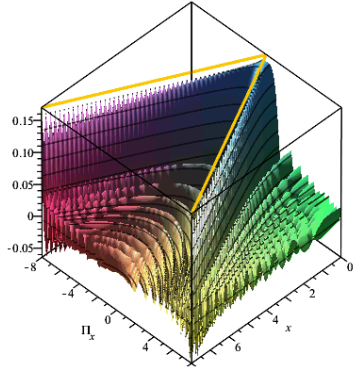}
  \caption{The Wigner function of cosmic string filled universe. The corresponding classical trajectory is
denoted by goldline loci. This figure is
plotted for $\Omega_{cs}=1$.}
\end{figure}
\section{Conclusion}
In this paper we studied the deformation  or
phase space quantization of a flat quantum FLRW
model, whose matter is either a fluids of radiation plus dust, or cosmic string.
We show that the peaks of Wigner quasiprobability distribution function of quantum universe filled with cosmic string are coincide with the emerged universes whose are filled with cosmic string and radiation fluids.
Also, for a universe filled by the radiation and dust fluids, the peaks of Wigner quasiprobability distribution function are coincide with the same classical universes but with modified density parameter of radiation.
Consequently the behaviour of emerged quantum universes are different with the corresponding classical models for the very small  values of scale factors, where the quantum universes are non-singular. On the other hand, for  large values of scale factor, the emerged quantum universes are coincide with corresponding classical universes. This behavior can be
interpreted that the classical approximation of the Universe becomes better
and better as the universe expands.
We believe the above results offer an insight into the relation between classical and quantum cosmologies and that the particular simple models that we have studied may serve as useful starting point for more ambitious investigation.
{ We are aware that our
results are obtained within a very simple as well as restricted
setting. Nevertheless, we think they are intriguing and provide
motivation for subsequent research works. A wider analysis, with
less restrictive cosmologies  and/or other matter fields, should follow.
The presence of phenomenological fluid matter  was
broadly used in, e.g., \cite{Matter} so that exact solutions of
the (simplified) WDW equation could be obtained. Using instead, e.g., a scalar , spinor or Yang-Mills fields
would be more generic and more realistic from the
point of view of matter interaction with the gravitational
field in a high energy regime, where
quantum effects can be expected. We are leaving the above
enticing research lines for future works.
}

\bibliographystyle{elsarticle-num}

\end{document}